\documentstyle[epsf,12pt,a4]{article}

\newtheorem{theorem}{Theorem}
\newtheorem{prop}{Proposition}
\newtheorem{definition}{Definition}

\begin{document}

\def\r{{\bf r}} \def\v{{\bf v}} \def\x{{\bf x}} \def\R{{\bf R}}
\def\Z{{\bf Z}} \def\del{\partial} \def\Lap{\bigtriangleup}
\def\Div{{\rm div}\ } \def\rot{{\rm rot}\ } \def\curl{{\rm curl}\ }
\def\grad{{\rm grad}\ } \def\Tr{{\rm Tr}} \def\^{\wedge} \def\real{{\rm
    re}} \def\image{{\rm im}} \def\goinf{\rightarrow\infty}
\def\goes{\rightarrow} \def\bm{\boldmath} \def\-{{-1}} \def\inv{^{-1}}
\def\sqr{^{1/2}} \def\isqr{^{-1/2}}

\def\eqna#1{\begin{eqnarray} #1 \end{eqnarray}}
\def\reff#1{(\ref{#1})}
\def\vb#1{{\partial \over \partial #1}} 
\def\Del#1#2{{\partial #1 \over \partial #2}}
\def\Dell#1#2{{\partial^2 #1 \over \partial {#2}^2}}
\def\Dif#1#2{{d #1 \over d #2}}
\def\Lie#1{ {\cal L}_{#1} }
\def\diag#1{{\rm diag}(#1)}
\def\abs#1{\left | #1 \right |}
\def\rcp#1{{1\over #1}}
\def\paren#1{\left( #1 \right)}
\def\brace#1{\left\{ #1 \right\}}
\def\bra#1{\left[ #1 \right]}
\def\angl#1{\left\langle #1 \right\rangle}
\def\lvector#1#2#3#4{\paren{\begin{array}{c} #1
         \\ #2 \\ #3 \\ #4 \end{array}}}
\def\vector#1#2#3{\paren{\begin{array}{c} #1 \\ #2 \\ #3 \end{array}}}
\def\svector#1#2{\paren{\begin{array}{c} #1 \\ #2 \end{array}}}
\def\matrix#1#2#3#4#5#6#7#8#9{
        \left( \begin{array}{ccc}
                        #1 & #2 & #3 \\ #4 & #5 & #6 \\ #7 & #8 & #9
        \end{array}     \right) }
\def\smatrix#1#2#3#4{
        \left( \begin{array}{cc} #1 & #2 \\
                #3 & #4 \end{array}     \right) }
\def\GL#1{{\rm GL}(#1)}
\def\SL#1{{\rm SL}(#1)}
\def\PSL#1{{\rm PSL}(#1)}
\def\O#1{{\rm O}(#1)}
\def\SO#1{{\rm SO}(#1)}
\def\IO#1{{\rm IO}(#1)}
\def\ISO#1{{\rm ISO}(#1)}
\def\U#1{{\rm U}(#1)}
\def\SU#1{{\rm SU}(#1)}

\def\Teich{{Teichm\"{u}ller }}
\def\Poin{{Poincar\'{e} }}
\def\Gam{\mbox{$\Gamma$}}
\def\a{{\bf a}}
\def\d{{\rm d}}
\def\n{\mbox{$n$}}
\def\hh{{h}}
\def\g{{\rm g}}
\def\Bsvn{VII${}_0$ }
\def\ho{\overline{h}}
\def\eo{\overline{{\rm e}}}
\def\ao{\overline{\alpha}}
\def\tR{{}^{(3)}\! R}
\def\fR{{}^{(4)}\! R}
\def\SR{\mbox{$S^2\times E^1$} }
\def\Isom{{\rm Isom}}
\def\Esom{{\rm Esom}}
\def\SL{\widetilde{{\rm SL}}(2,\R)}
\def\SLx{{\rm SL}(2,\R)}
\def\Nil{{\rm Nil}}
\def\Sol{{\rm Sol}}
\def\Lapp{\bar g}

\def\uh#1#2{\hh^{#1#2}}
\def\dh#1#2{\hh_{#1#2}}
\def\ug#1#2{\g^{#1#2}}
\def\dg#1#2{\g_{#1#2}}
\def\uug#1#2{\tilde{\g}^{#1#2}}
\def\udg#1#2{\tilde{\g}_{#1#2}}
\def\udh#1#2{\tilde{\hh}_{#1#2}}
\def\ustd#1#2{\tilde{h}_{#1#2}^{\rm std}}

\def\BVII#1{\mbox{Bianchi VII${}_#1$} }
\def\BVI#1{\mbox{Bianchi VI${}_#1$} }
\def\four#1{{}^{(4)}\! #1}
\def\sz{\mbox{VI${}_0$}}
\def\svz{\mbox{VII${}_0$}}

\def\diffeos{diffeomorphisms}
\def\diffeo{diffeomorphism}
\def\MM{\four{M}}
\def\uMM{\four{\tilde{M}}}
\def\Mtil{\tilde M}

\def\Mt{\mbox{$M_t$}}
\def\uMt{\mbox{$\tilde{M}_t$}}
\def\uMi{\mbox{$\tilde{M}_{t_0}$}}
\def\UC{\mbox{$(\uMM,\udg ab)$}}
\def\CU{\mbox{$(\MM,\dg ab)$}}
\def\UCh{\mbox{$(\uMt,\udh ab)$}}
\def\UCs{\mbox{$(\Mtil,\udh ab)$}}
\def\UChi{\mbox{$(\uMti,\udh ab)$}}
\def\CUh{\mbox{$(\Mt,\dh ab)$}}
\def\Isomf{\mbox{$\Esom\uMt$}}
\def\Isomff{\mbox{$\Esom(\uMt,\uMM)$}}
\def\teich{{\bf \tau}}
\def\dataset{$(\dh ab, K_{ab})$}
\def\onedata{$(\dh \mu\nu, K_{\mu\nu})$}

\def\s#1{\sigma^{#1}}
\def\a#1#2{a_{#1}{}^{#2}}
\def\dug#1#2{g_{#1}{}^{#2}}
\def\h#1{\hh_{#1#1}}
\def\H#1#2{H_{#1#2}}
\def\f#1#2{f^{#1}{}_{#2}}

\def\wa{&=&}
\def\wb{&\equiv&}

\def\I{I}

\def\abstract#1{\begin{center}{\bf ABSTRACT}\end{center}
\par #1}
\def\title#1{\begin{center}{\large {#1}}\end{center}}
\def\author#1{\begin{center}{\sc #1}\end{center}}
\def\address#1{\begin{center}{\it #1}\end{center}}

\def\pubnum{322/COSMO-68}

\begin{titlepage}
\hfill
\parbox{6cm}{{TIT/HEP-\pubnum} \par 1996}
\par
\vspace{1.5cm}
\begin{center}
\Large
Dynamics of compact homogeneous universes
\end{center}
\vskip 1.3cm
\author{Masayuki TANIMOTO ${}^{*}$,
Tatsuhiko KOIKE  ${}^{**}$,
and Akio HOSOYA  ${}^{***}$
}
\address{Department of Physics, Tokyo Institute of \\ Technology,
Oh-Okayama 2-12-1, Meguro-ku, Tokyo 152, Japan}
\vskip 1 cm

\abstract{ A complete description of dynamics of compact locally
  homogeneous universes is given, which, in particular, includes
  explicit calculations of \Teich deformations and careful counting of
  dynamical degrees of freedom.  We regard each of the universes as a
  simply connected four dimensional spacetime with identifications by
  the action of a discrete subgroup of the isometry group.  We then
  reduce the identifications defined by the spacetime isometries to ones
  in a homogeneous section, and find a condition that such spatial
  identifications must satisfy.  This is essential for explicit
  construction of compact homogenoeus universes.  Some examples are
  demonstrated for Bianchi II, VI${}_0$, VII${}_0$, and I universal
  covers.  }


\end{titlepage}

\addtocounter{page}{1}

\section{Introduction}

In relativistic and observational cosmology, we often use a simplified
spacetime model having restricted dynamical degrees of freedom. In
particular, the well-known homogeneous and isotropic (FRW) models
\cite{Wein}, in which the spatial sections are assumed to be homogeneous
and isotropic, have been successful.  On the other hand, a wider class
of models, known as the Bianchi homogeneous models \cite{LL,Ryan,Wa}, in
which the spatial sections are assumed to be homogeneous but not
isotropic, have been largely used in relativity and quantum cosmology
\cite{Ryan,BKL,Hal}.  In the models except type IX, each spatial section
has been regarded as {\it open}.  The open topology, however, is not a
sole possibility.  For example, the ``open'' model in the homogeneous
and isotropic models, which has constant negative spatial curvature, and
belongs to Bianchi type V, can be regarded to be spatially compact if
spatial points are appropriately identified with some discrete subgroup
of the isometry group.  The purpose of this paper is, in fact, to
investigate a class of the models in which each spatial section has {\it
  compact} topology.  The compactness of space is physically reasonable
due to its finite spatial volume.

The crucial point of the arguments in this paper is that the compactness
of locally homogeneous space, in general, brings about new degrees of
freedom of deformations, known in mathematics as \Teich deformations.
They preserve the local geometry but change the global one.  This can be
easily understood if we regard the \Teich deformations of a compact
locally homogeneous space as a homogeneous space (i.e., a covering
space) with varying identifications.  A space spanned by independent
\Teich deformations, and its coordinates are referred to as the {\it
  \Teich space} and the {\it \Teich parameters}, respectively.  We shall
shed light on these degrees of freedom of the \Teich deformations, which
in fact have been often disregarded so far.  The field of (2+1)-gravity
\cite{HNCa} was exceptional, but we shall take a somewhat different
approach.  The \Teich deformations are of great interest, because they
would carry part of the dynamical degrees of freedom.  We shall give a
complete framework to do a concrete analysis concerning the \Teich
deformations.  We also carefully count the total dynamical degrees of
freedom.

In Ref.\cite{KTH}, referred to as \I\ hereafter, we presented a
treatment of three-dimensional compact homogeneous Riemannian manifolds,
which will be a basis of our arguments in the present work.  We, there,
(1) gave the possible eight types (a$\sim$h) of homogeneous universal
covers, which are closely related to Thurston's eight
geometries\cite{Th,Sc}, (2) classified compact quotients (a1/1, b/1,
etc.), and (3) gave \Teich spaces by explicitly finding embeddings of
covering groups in the isometry groups of the universal covers, which
enable us to perform explicit calculations.  To investigate the dynamics
of compact homogeneous universes, we must first show how we can adapt
such knowledge of compact homogeneous three-manifolds in the context of
relativity in four dimensions.

Our strategy to this will be as follows.  We begin with considering a
four-dimensional universal cover \UC, which is a simply connected
Lorentzian manifold, and then take identifications in \UC\ so as to make
each 3-surface \UCh\ compact.  To utilize our knowledge about compact
homogeneous 3-manifolds, we translate the identifications in \UC\ into
those in \UCh.  We find that the identifications must be {\it extendible
  isometries} of \UCh, which have natural extention in \UC.  We make
\UCh\ compact by the action of a discrete subgroup \Gam\ of the group
$\Isomf\subset\Isom\uMt$ of extendible isometries, where $\Isom\uMt$ is
the isometry group of \UCh.  Once given $\Gam\subset\Isomf$ on \UCh, all
necessary identifications in \UC\ are automatically determined by the
natural extension of \Gam.  The quotient \CU\ is a solution of a local
equation, e.g., Einstein's equation, if and only if \UC\ is a solution
of the same equation.

The organization of this paper is as follows.  In section \ref{prelim},
we first give the definition of \Teich deformations, and then briefly
review the classification of compact homogeneous 3-manifolds given in
\I.  In section \ref{mi}, we establish the prescription for
identifications, and discuss possible four-dimensional universal covers.
We also discuss how we can eliminate the ``gauge'' degrees of freedom,
and thereby we give how to find the dynamical degrees of freedom.  In
section \ref{tt}, we apply the framework of the previous section to
concrete models.  There, we give the time-development of the \Teich
parameters, establish the dynamical variables and give the number of the
dynamical degrees of freedom for each case.  The final section is
devoted to conclusions.  We employ the abstract index notation (See e.g.
\cite{Wa}) throughout this paper.

\section{Preliminaries}
\label{prelim}

We give definitions of \Teich deformations in the first subsection.  In
the next subsection, we briefly sketch the classification scheme, given
in \I, of compact homogeneous 3-manifolds, though we will not duplicate
the results of the classification.  This will be helpful for the
subsequent discussions.  For the explicit results, see tables 1 and 2
\cite{er} in \I, and section V of \I.  We consider only complete
Riemannian manifolds, and shall drop the word ``complete'' hereafter.

\subsection{\Teich deformations}
\label{candt}

\def\preUC{\mbox{$(\tilde M,\tilde h{}_{ab})$}} We define \Teich
deformations of a Riemannian manifold $(M,\dh ab)$ as follows.
\begin{definition}[\Teich deformations]
  Let $(M,\dh ab)$ be a Riemannian manifold.  Then, smooth and
  non-isometric deformations of $\dh ab$ are called \Teich deformations
  if they leave the universal cover \preUC\ globally conformally
  isometric.
\end{definition}
In this definition, a globally conformal isometry means a conformal
isometry with {\it constant} conformal factor.  For definition of
coverings, see, e.g., Ref.\cite{ST}.

In other words, \Teich deformations are deformations induced by
variations of a covering group $\Gamma$ of the universal cover \preUC.
Here, a covering group \Gam\ is a representation (or an embedding) of
the fundamental group $\pi_1M$ into $\Isom\tilde M$, the isometry group
of \preUC, where $M$ can be realised as $\tilde M/\Gam$.  We denote the
space of all covering groups as ${\rm Rep}(\pi_1M,\Isom\tilde M)$.  Note
that not all variations of $\Gamma$ correspond to independent \Teich
deformations.  In fact, two Riemannian manifolds $\tilde M/\Gamma'$ and
$\tilde M/\Gamma$ are isometric if
\begin{equation}
        \Gamma'= a\circ\Gamma\circ a\inv
        \label{pre-2}
\end{equation}
holds for an isometry $a\in \Isom\tilde M$.  $\Gamma'$ is called the
{\it conjugation} of \Gam\ by $a$.  If we denote the equivalence
relation by conjugation as $\sim$, the \Teich space, ${\rm
  Teich}(M,\tilde h_{ab})$, for $M$ and $\tilde h_{ab}$, is defined as
\begin{equation}
        {\rm Teich}(M,\tilde h_{ab})={\rm Rep}(\pi_1M,\Isom\tilde M)/\sim.
        \label{Teich-space}
\end{equation}

Although in this paper we are interested only in locally homogeneous
metrics, it is worth noting that the definition of \Teich deformations
here concerns wider classes of metrics.  For example, even though the
universal cover admits only one Killing vector, the quotients can
smoothly deform if we smoothly vary the identifications along the
Killing orbit.

\subsection{The classification of compact homogeneous 3-manifolds}
\label{cl}

We briefly sketch the classification scheme, given in \I, of compact
homogeneous 3-manifolds.

Consider a pair $(M,G)$ of a manifold $M$ and a group $G$ acting
transitively on $M$ with compact isotropy subgroup.  Note that we can
construct a homogeneous manifold $(M,\dh ab)$ by first giving an
arbitrary metric at a point $p$ in $M$, averaging it by the isotropy
subgroup, then finally sending it by the actions of $G$.  Since $G$ is
transitive, the resulting metric $\dh ab$ is guaranteed to be
homogeneous.  The isometry group of $(M,\dh ab)$ would be isomorphic to
$G$, or contain $G$ as a subgroup of the isometry group.  Note that if
we give various metrics at $p$, then we obtain many homogeneous metrics
on $M$.  Hence the pair $(M,G)$ can be considered as an equivalence
class of homogeneous manifolds whose isometry groups are isomorphic to
$G$, or contain $G$ as a subgroup of the isometry groups.  Such a pair
$(M,G)$ is called a {\it geometry}.  If two geometries $(M,G)$ and
$(M,G')$ have an inclusion relation $G\subset G'$, then $(M,G)$ is
called a {\it subgeometry} of $(M,G')$.  If geometry $(M,G)$ is not a
subgeometry of any geometry, then $(M,G)$ is called a {\it maximal}
geometry, and if geometry $(M,G)$ does not have any subgeometry, then
$(M,G)$ is called a {\it minimal} geometry.

Our starting point of classification of compact homogeneous manifolds is
following Thurston's theorem \cite{Th}.
\begin{theorem}
  Any maximal, simply connected three-dimensional geometry which admits
  a compact quotient is equivalent to the geometry $(M,\Isom M)$ where
  $M$ is one of $E^3$, $H^3$, $S^3$, $S^2\times\R$, $H^2\times\R$,
  $\widetilde{SL}(2,\R)$, Nil, and Sol.
\end{theorem}
A brief proof and accounts of the eight geometries in the theorem are
found in Ref.\cite{Sc}.

Note that this theorem concerns only {\it maximal} geometries.  This
seems, however, too restricted for our purpose.  For example, while we
are interested in the closed FRW universe, of which a spatial section
corresponds to maximal geometry $(S^3,\SO4)$, we are also, and maybe
more, interested in the Bianchi IX universe, of which a spatial section
is subgeometry $(S^3,\SU2)$ of geometry $(S^3,\SO4)$.  Hence we should
concern all {\it non-maximal} geometries which admit compact quotients,
too.  The above theorem, however, is still of great use, because of the
following fact.  That is, the ``maximized'' geometry of any non-maximal
geometry admitting a compact quotient must admit a compact quotient,
because the group of the maximal geometry must contain the covering
group which makes the non-maximal geometry compact, and it must make the
maximal geometry compact.  This simple fact and Thurston's theorem lead
us to the investigation of all possible subgeometries of Thurston's
eight maximal geometries.  Of course, we must note that not all
subgeometries of the eight geometries admit compact quotients.  We need
to check explicitly that the subgeometry certainly admits a compact
quotient.

Although, as we have seen, the use of ``geometry'' is convenient to
carry out classification of homogeneous manifolds, it is useful to
switch to the conventional ``Riemannian manifold'' representation for
physical applications.  As we have noted, a geometry is an equivalence
class of homogeneous Riemannian manifolds.  Let $R$ be the set of all
homogeneous manifolds belonging to a geometry, and let $\bar R$ be the
quotient set of $R$ by all possible diffeomorphisms and globally
conformal transformations.  Our classification for all possible
universal covers admitting compact quotients is described in terms of
$\bar R$'s.  In \I, we labeled those $\bar R$'s as ``a1, a2, b, c, ...,
h''.  For each element of quotient $\bar R$, we choose a representative
element $(M,\dh ab)$, and call metric $\dh ab$ the {\it representative
  metric} or the {\it standard metric}, whose explicit form for each
type is also given in \I.

In getting the standard metrics, the Bianchi-Kantowski-Sachs-Nariai
(BKSN) classification \cite{Bi,Na,KS} is useful, which concerns all
minimal 3-geometries;
\begin{theorem}
  Any minimal, simply connected three-dimensional geometry is equivalent
  to $(M,G)$ where $M=\R^3$, $G=$one of Bianchi I to Bianchi VIII
  groups; $M=S^3$, $G=$Bianchi IX group; or $M=S^2\times\R$,
  $G=SO(3)\times\R$.
\end{theorem}
This is useful in that the invariant metrics for these groups are well
known.  These geometries are called the {\it BKSN} minimal geometries.
We take this opportunity to show the correspondence of Thurston's eight
geometries and the BKSN minimal geometries, which is shown in table 1.
(Such a correspondence was first pointed out by Fagundes \cite{Fa}
without referring to subgeometry.)

\begin{table}[htb]
\begin{center}
\begin{tabular}{|l|l|}
      \hline
        Thurston's maximal geometries & BKSN minimal geometries \\
	 \hline\hline
        $(E^3,\Isom E^3)$ & $(\R^3,{\rm BI})$, $(\R^3,{\rm BVII_0})$ \\
	 \hline
        $(H^3,\Isom H^3)$ & $(\R^3,{\rm BV})$, $(\R^3,{\rm BVII}_{\it a})$
	\\ \hline
        $(S^3,\Isom S^3)$ & $(S^3,{\rm BIX})$ \\ \hline
        $(S^2\times\R,\Isom S^2\times\R)$ & $(S^2\times\R,{\rm KSN})$ \\
	 \hline
        $(H^2\times\R,\Isom H^2\times\R)$ & $(\R^3,{\rm BIII})$ \\ \hline
        $(\SL,\Isom \SL)$ & $(\R^3,{\rm BVIII})$, $(\R^3,{\rm BIII})$ \\
	 \hline
        $(Nil,\Isom Nil)$ & $(\R^3,{\rm BII})$ \\ \hline
        $(Sol,\Isom Sol)$ & $(\R^3,{\rm BVI_0})$ \\ \hline
\end{tabular}
\end{center}
        \caption{The correspondence between Thurston's maximal geometries
        and BKSN minimal geometries.
        BI is an abbreviation for Bianchi I group, and similar for
        BII, BIII, etc.
        KSN is equivalent to $\Isom S^2\times\Isom\R$.
        The missing Bianchi types IV and VI${}_a$ do not admit
        compact quotients. \label{tab:1}}
\end{table}

It should be noted that an $\bar R$ does not always correspond to a
single geometry, though the converse is true.  For example, consider
Bianchi I minimal geometry $(\R^3,\R^3)$, where the left $\R^3$ stands
for the manifold homeomorphic to $\R^3$, while the right $\R^3$ stands
for the 3-dimensional translation group acting on the manifold $\R^3$.
Then the corresponding $R$ is the set of Riemannian manifolds $(\R^3,\dh
ab)$'s, where $\dh ab=\dh\mu\nu(\d x^\mu)_a(\d x^\nu)_b$ with
$\dh\mu\nu$ being positive definite symmetric $3\times3$ matrix.  All
such $(\R^3,\dh ab)$'s, however, are isometric to $(\R^3,\eta_{ab})$,
where $\eta_{ab}$ is the standard Euclid metric
$\eta_{ab}=\delta_{\mu\nu}(\d x^\mu)_a(\d x^\nu)_b$.  We thus have $\bar
R=\{ (\R^3,\eta_{ab})\}$.  On the other hand, it is manifest that
maximal geometry $(\R^3,\IO3)$ also gives rise to the same $\bar R$,
since $\IO3$ is the isometries of $\eta_{ab}$.  (The $\bar R$ of this
example is classified to type a2 in our classification.  Similarly,
Bianchi types II, VII${}_0$, and VI${}_0$ give rise, respectively, to
universal cover types b, a1, and f, on which compact models will be
discussed specifically in subsequent sections.)

Let us summarize the actual procedure for our classification.  First,
list up Thurston's eight geometries and all their subgeometries.
Enumerate all possible compact quotients of Thurston's eight geometries,
and check whether the subgeometries admit them.  Find out $\bar R$'s for
the subgeometries which admit a compact quotient.  Note that this
includes the explicit determinations of the standard metrics.  This
completes the classification of the universal covers which admit a
compact quotient.  The classification of the compact quotients that
those universal covers admit can be carried out by using the results in
Ref.\cite{Wo,Oh,KLR}.  The \Teich parameters are usually defined as
components of the identification generators acting on the standard
metric.  Our classification, as a result, consists of the classification
of universal covers, the classification of compact quotients, and
explicit parametrizations of the \Teich spaces.

\section{The framework of construction}
\label{mi}

In the first subsection, we show a method of construction of compact
homogeneous universes, and then in the second subsection, we discuss the
dynamical degrees of freedom of a system of compact homogeneous
universes.  Note, however, that the second subsection is not independent
of the first one.  Our construction of a system of compact homogeneous
universes is completed by the discussions there.

\subsection{Universal covers and identifications}
\label{ui}

We mean by a compact homogeneous universe a smooth Lorentzian 4-manifold
\CU\ which admits a foliation by compact homogeneous spatial leaves
(sections), and denote the universal cover of \CU\ as \UC.  It is
important that \CU\ inherits all the local properties from \UC by a
covering map.  Hence, we can think that the local and the global
geometries are carried by the universal cover \UC\ and the covering map,
respectively, and thus we can look into them separately.  First, we
shall consider how we can take the covering map, or ``identifications'',
when a universal cover \UC\ is given.  After that, we shall consider
what universal covers are appropriate for our purpose.

The identifications in \UC\ act on each homogeneous 3-section \UCh\ of
$t=$const., where $t$ parametrizes the homogeneous sections of \UC.  For
simplicity, we omit the argument $t$ of the metric $\tilde h_{ab}$ as
far as no confusions occur.  Let $\Isom\uMt$ be the isometry group of
\UCh.  It is very important to note that we {\it cannot} make, in
general, the homogeneous 3-manifold \UCh\ compact by the action of a
discrete subgroup of $\Isom\uMt$, since for \CU\ to be a smooth
Lorentzian manifold, the covering group, \Gam, of the section needs to
preserve the extrinsic curvature, as well as the spatial metric.  To
give a suitable prescription of compactification, we convert this
requirement of the smoothness of \CU\ into the following statement;
Since \CU\ is obtained by taking identifications in \UC, for any two
points which are identified, there should exist an isometry of \UC\ 
({\it not} of \UCh) which maps one to the other.  Hence, if we define
the {\it extendible isometry group} of \UCh, $\Isomf\subset\Isom\uMt$,
as below, then we obtain a complete prescription for construction of a
compact homogeneous universe, as shown subsequently;

\begin{definition}[Extendible isometry group]
  Let \UCh\ be a spatial section of \UC.  An extendible isometry is the
  restriction on $\uMt$ of an isometry of \UC\ which preserves $\uMt$.
  They form a subgroup of $\Isom\uMt$.  We call it the extendible
  isometry group, and denote it as \Isomff, or simply \Isomf.
  Obviously, an extendible isometry $a\in\Isomf$ has the natural
  extension on $\uMM$ which is an element of $\Isom\uMM$ and preserves
  $\uMt$.  We call such the natural extension on $\uMM$ the extended
  isometry of $a$, or simply the extension of $a$.
\end{definition}
\begin{prop}
        \label{prop1}
        The identifications on an initial surface \UCh\ must be
        implemented in \Isomff,
\begin{equation}
        \Gam\subset\Isomff,
        \label{mc10}
\end{equation}
to get a compact homogeneous universe out of a given four-dimensional
universal cover \UC.  Moreover, the identifications acting on whole \UC\ 
are determined by the action of the extension of $\Gam$ on $\uMM$.
\end{prop}

For example, Kasner type metric,
\begin{equation}
        \d s^2=-\d t^2+t^{2p_1}\d x^2+t^{2p_2}\d y^2+t^{2p_3}\d z^2,
        \label{mc5}
\end{equation}
where $p_1\sim p_3$ are constants, has Euclid spaces as its spatial
part.  Suppose \UCh\ is such a Euclid space, where the metric is given
by
\begin{equation}
        \d l^2=t^{2p_1}\d x^2+t^{2p_2}\d y^2+t^{2p_3}\d z^2.
        \label{mc7}
\end{equation}
As long as a generic case is concerned where $p_\alpha$'s are all
different, the (continuous) isometries of the 4-metric \reff{mc5} are
only the translations generated by $\del/\del x$, $\del/\del y$, and
$\del/\del z$, which form group $\R^3$.  Since they all preserve the
spatial sections, we find $\Esom_0\uMt\simeq\R^3$.  ({$\Esom_0\uMt$ is
  the identity component of \Isomf, and similar for $\Isom_0\uMt$.}) On
the other hand, rotations generated by vector
\begin{equation}
        k_3\equiv -t^{-(p_1-p_2)}y\Del{}{x}+t^{p_1-p_2}x\Del{}{y},
        \label{mc8}
\end{equation}
and the similar vectors obtained by permutations of indices also form
intrinsic isometry group, $\SO3$, of \UCh.  This shows
$\Esom_0\uMt\neq\Isom_0\uMt\simeq \ISO3$.  On a homogeneous section
\UCh, if we choose a covering group \Gam\ in \Isomf\ such as to make
\UCh\ compact, we obtain a compact homogeneous universe as a
four-dimensional manifold through proposition \ref{prop1}.

We now discuss what universal covers are appropriate for our purpose.
If the topology of $\MM$ is fixed, then manifold $\uMM$ is uniquely
determined.  Hence we only need to consider possible metrics, $\udg
ab$'s.  Let us consider the case where each homogeneous spatial section
\UCh\ corresponds to a Bianchi minimal geometry.  It is also
straightforward to adapt the following description for the KSN minimal
geometry.  By the definition of our compact homogeneous universes, the
metric should be of the form
\begin{equation}
        \d s^2= -N^2(t,\x)\d t^2+\dh\alpha\beta(t)
                (N^\alpha(t,\x)\d t+\s\alpha)(N^\beta(t,\x)\d t+\s\beta),
        \label{mc1}
\end{equation}
where $(t,\x)$ are local coordinates, $\s\alpha$ the invariant 1-forms,
and $\alpha,\beta,\ldots$ run from 1 to 3.  The spatial metric
$\dh\alpha\beta(t)\s\alpha\s\beta$ is, indeed, homogeneous on each
section $t=$const..

For a metric with generic lapse and shift functions, the extendible
isometry group \Isomf\ contains only the identity element so that we
cannot use the prescription for the compactification.  One might argue
that there would remain possibility to find a discrete group $\Gam'$ for
the compactification if the lapse and shift functions have some
periodicity.  However, such a discrete group does not contain continuous
parameters.  Since we are interested in the case of maximal number of
continuous parameters in the initial identifications, we demand that
\Isomf\ is transitive and therefore the lapse and shift functions are
independent of the spatial coordinates $\x$;
\begin{equation}
        \d s^2= -N^2(t)\d t^2+\dh\alpha\beta(t)
                (N^\alpha(t)\d t+\s\alpha)(N^\beta(t)\d t+\s\beta).
        \label{mc2}
\end{equation}

This metric becomes the following form
\begin{equation}
        \d s^2= -\d t^2+\dh\alpha\beta(t)\s\alpha\s\beta,
        \label{mc3}
\end{equation}
by the induced map of a \diffeo\ which preserves each homogeneous
section of $t={\rm const.}$.  We shall mainly focus on this type of
metrics hereafter.

We here comment on an intuitive prescription for identifications, which
is also useful particularly for the metric \reff{mc3}.  Note that the
normal geodesics emerging from a section \uMt\ are uniquely defined,
provided that they are parametrized by proper time $\tau$.  We refer to
the exponential map $\exp\tau n^a(t)$ which is defined with respect to
the normal vector field $n^a(t)$ on \uMt\ as the {\it normal map}.
({The image of \uMt\ by the normal map $\exp\tau n^a(t)$ is {\it not}
  generally $\uMt{}_{{}'}$ for some $t'$ when considering metric
  \reff{mc1}.  For metric \reff{mc3}, we of course have $\exp\tau
  n^a(t):\, \uMt\goes\uMt{}_{+\tau}$.}) Obviously, if two points, $a$
and $b$, on $\uMt$ are identified, any two points mapped by the normal
map should continue to be identified,
\begin{equation}
        \forall \tau\in\R;\; (\exp\tau n^a(t))(a)\sim(\exp\tau n^a(t))(b).
        \label{eq:normal}
\end{equation}
Hence, we can determine how the identifications evolve in time, in terms
of geodesics in a given four-dimensional universal cover.  For the
metric \reff{mc3}, since the hypersurface-orthogonal geodesics coincide
with the t-axes, we immediately obtain the following useful proposition.
\begin{prop}
        \label{prop2}
        In terms of the coordinates $(t,\x)$ of metric \reff{mc3}, if at
        the initial surface $t=t_0$ an identification is specified as
        $(t_0,\x)\sim(t_0,a\x)$, where $a$ is a free action on the
        coordinate space, then at any time $t$ we must have
        $(t,\x)\sim(t,a\x)$.
\end{prop}
That is, if we take identifications in \Isomf\ on an initial surface of
the metric \reff{mc3}, and describe them in terms of the spatial
coordinates $\x$, then the description of the identifications holds for
any time $t$.

By proposition \ref{prop2}, one might conclude that no interesting
global, i.e. Teichm\"{u}ller, deformations occur for the metric
\reff{mc3}, since the identifications on each homogeneous section in
terms of the spatial coordinates do not vary with time.  However, this
is not the case.  Remember that the \Teich deformations are defined with
respect to the intrinsic geometries of the three dimensional sections
\CUh.  Variation of metric with time does cause \Teich deformations with
time, and if there exists difference between $\Isomf$ and $\Isom\uMt$,
as in the example below proposition \ref{prop1}, the situation becomes
much richer.  We will comment on this point again at the end of the next
subsection.

\subsection{Dynamical degrees of freedom}
\label{degree}

Let us consider a universe characterized by an initial data set
\dataset, where $\dh ab$ and $K_{ab}$ are the spatial metric and the
extrinsic curvature of the initial spatial section $M$.  To give a
possible \dataset\ in the case that $M$ has nontrivial topology and
\dataset\ is locally homogeneous, we first need to cover $M$ with some
open patchs, define coordinates for each patch, and give a
transformation function for each overlap of two patchs.  That is, we
need to set an atlas.  After that, if we give an \onedata\ on a point
with respect to the coordinates defined in a patch, we can send \onedata
\cite{ex} to all points in the patch by the transitive group action, and
finally assign the values of the data set to all patchs by the
transformation functions and the group actions.  Hence the information
of the data \dataset\ is equivalent to the information of the value of
\onedata\ on a point {\it and} the way of taking transformation
functions if patchs are fixed.  However, it can be easily imagined that
it is very difficult to count the number of possible \onedata's and the
number of possible ways of taking transformation functions up to
diffeomorphism.  We can accomplish this counting, using coverings, as
follows.

As stated in the previous subsection, we think that a set, $U$, of
universal covers, \UC's, carries the degrees of freedom of local
geometry like local curvatures, and the covering maps do the degrees of
freedom of global geometry like \Teich parameters.  In this approach, it
is evident that we need to eliminate the degrees of freedom of all
possible \diffeos\ in $U$.  We introduce the equivalence relation in $U$
by \diffeos, and denote the resulting set of universal covers as $\bar
U$.  If we fix a homogeneous surface \UCh\ in a $u\equiv\UC\in \bar U$
\cite{id}, and suppose $\Gam\subset\Isomf$ makes \UCh\ compact, then we
can identify a pair $(u,\Gam)$ with a compact homogeneous universe
through proposition \ref{prop1}.  We denote the set of $(u,\Gam)$'s of
all possible \Gam's for a fixed $u$ as $C_u$.  If $\Gam'=
\phi\circ\Gam\circ\phi\inv$ holds for a $\phi\in\Isomf$, then the
resulting compact homogeneous universe, $(u,\Gam')$, is isometric to
$(u,\Gam)$.  In this sense, the freedom of taking conjugations of \Gam\ 
by \Isomf\ also corresponds to ``gauge'' freedom (cf. Sec.\ref{candt}).
Introducing the equivalence relation in $C_u$ by the conjugations, we
get the quotient set $\bar C_u$.  Now, our dynamical system, $\bar C$,
of compact homogeneous universes is equivalent to the set $\bar
C\equiv\{c|\; c\in\bar C_u,\; u\in\bar U\}$.

If we choose a homogeneous section arbitrarily for each element in $\bar
C$, we will have a set, $\bar I$, of initial data sets, \dataset's, on a
fixed compact 3-manifold.  In $\bar I$, there are no elements which are
isometric to each other, since for any different elements in $\bar C$
are non-isometric, and the development of an initial data set is unique.
The set $\bar I$ is therefore equivalent to the set we considered at the
beginning of the subsection.

The following proposition is now trivial.
\begin{prop}
        \label{prop1-1}
        The number, $\dim\bar C$, of degrees of freedom of a system of
        compact homogeneous universes is the sum of the number,
        $\dim\bar U$, of the degrees of freedom of the four-dimensional
        universal covers up to isometry, and the number, $\dim\bar C_u$,
        of degrees of freedom of initial identifications, i.e.  covering
        groups on an initial section, up to conjugations taken by the
        extendible isometry group.
\end{prop}
Hereafter, by a construction of compact homogeneous universes, we mean a
construction with explicit determination of representatives of the
universes in the above sense, so that the number of arbitrary parameters
in a universal cover should be $\dim\bar U$, and the number of arbitrary
parameters in the identification generators should be $\dim\bar C_u$.
Note that for vacuum solutions of Einstein's equation for Bianchi class
A \cite{EM}(i.e. types I, II, VI${}_0$, VII${}_0$, VIII, and IX) and
type V, the metric components $\dh\alpha\beta(t)$ in Eq.\reff{mc3} is
``diagonalizable'', i.e. becomes diagonal by diffeomorphisms.  Hence we
will begin with the diagonal form metric to construct compact
homogeneous universes on Bianchi class A or type V.

Note that we take conjugations for \UCh\ only by \Isomf\ to obtain the
initial identification parameters as stated in proposition \ref{prop1},
while the \Teich parameters are defined with respect to conjugations by
full $\Isom\uMt$ (cf. Sec.\ref{candt}).  Roughly speaking, the
difference between the freedom of \Isomf\ and that of $\Isom\uMt$
corresponds to the freedom of giving initial `velocities' of \Teich
parameters, as we will see more explicitly through the examples in the
next section.

\section{Four compact homogeneous universe models}
\label{tt}

In this section, we construct four explicit models of compact
homogeneous universe, the b/1, f1/1, a1/1, and a2/1 models.  For each
model, we count the number of dynamical degrees of freedom and give the
time-development of the \Teich parameters.

To get the \Teich parameters of a compact section \CUh, we need to
compare two mathematical representations, i.e., \UCh\ with the covering
group \Gam, and the standard universal cover $(\Mtil,\tilde h_{ab}^{\rm
  std})$ with the covering group, $A$, parametrized by the \Teich
parameters.  ({In \I, the standard metrics were called the
  representative metrics.}) \Gam\ and $A$ are generated by the same
number, $n$, of generators, $\{g_i\}$ and $\{a_i\}$ ($i=1,\cdots,n$),
respectively.  $\{g_i\}$ and $\{a_i\}$ satisfy the same multiplication
rule of an extendible isometry group.  We can get the \Teich parameters
by finding the automorphism of $\Esom\tilde M$ which relates the two
sets of generators.  We shall do this first for the b/1 model, where we
will see the most {\it typical} calculation to get \Teich parameters.
Then the f1/1 and a1/1 models follow.  Finally, for the a2/1 model, we
present a different method in getting the time-development of the \Teich
parameters.

Our universal cover metrics are synchronous (Eq.\reff{mc3}) and diagonal
(See Sec. \ref{degree}).

\subsection{The b/1 model: a compact model on Bianchi II geometry}
\label{b}

\def\pfaci{\sqrt{\frac{\h3}{\h2}}} \def\pfacii{\sqrt{\frac{\h3}{\h1}}}
\def\pfaciii{\frac{\h3}{\sqrt{\h1\h2}}}

We start with the multiplication rule of Nil (=Bianchi II group);
\begin{equation}
        \vector{g^1}{g^2}{g^3}\vector{h^1}{h^2}{h^3}=
        \vector{g^1+h^1}{g^2+h^2}{g^3+h^3+g^1h^2},
        \label{b2-1}
\end{equation}
where $g,h\in {\rm Nil}$, and we shall use superscripts to denote the
components of a group element.  We use the same components
$(x^1,x^2,x^3)\equiv(x,y,z)$ as coordinates of $\uMt$.  The action of
Nil on $\uMt$ is defined by the left action on $(x,y,z)\in$Nil.  A
Nil-invariant (diagonal) metric is given by
\begin{equation}
        \d l^2=\h1\d x^2+\h2\d y^2+\h3(\d z-x\d y)^2,
        \label{b2-2}
\end{equation}
where $\h\alpha\,(\alpha=1\sim3)$ are constants, i.e. independent of
$(x,y,z)$.  The four dimensional universal cover metric of our concern
is of the form
\begin{equation}
        \d s^2=-\d t^2+\d l^2
        \label{b2-2'}
\end{equation}
with $\h\alpha$ being functions of $t$.  The vacuum solution is, of
course, known, but we proceed with calculation, leaving $\h\alpha$ free,
since they are complicated functions in the synchronous gauge and,
moreover, it enable us to apply the result also to models other than the
vacuum model.

We consider manifold ``b/1 $(n=1)$'', classified in \I, which is
probably the most stereotypical compact manifold modeled on Bianchi II
geometry.  The fundamental group $\pi_1$ is given by (See Eq.(118) in
\I)
\begin{equation}
        \pi_1=\angl{g_1,g_2,g_3;[g_1,g_2]g_3\inv,[g_1,g_3],[g_2,g_3]}.
        \label{b2-3}
\end{equation}
The topology of b/1 is illustrated in Fig.1.

To represent the generators of $\pi_1$, $g_i$'s, in $\Isomf={\rm Nil}$,
we put them as
\begin{equation}
        g_i=\vector{\dug i1}{\dug i2}{\dug i3},\, (i=1\sim3),
        \label{b2-4}
\end{equation}
and substitute these in the relations of $\pi_1$ (Eq.\reff{b2-3}).  We
then get the following;
\begin{equation}
        g_1=\vector{\dug 11}{\dug 12}{\dug 13},\,
        g_2=\vector{\dug 21}{\dug 22}{\dug 23},\,
        g_3=\vector{0}{0}{\Lapp_3{}^3},
        \label{b2-5}
\end{equation}
where $\Lapp_3{}^3\equiv\dug11\dug22-\dug12\dug21\neq0$.

\vspace{1cm}
\begin{figure}[btp]
\label{fig:0}
  \begin{center}
    \leavevmode
\epsfysize=5cm
\epsfbox{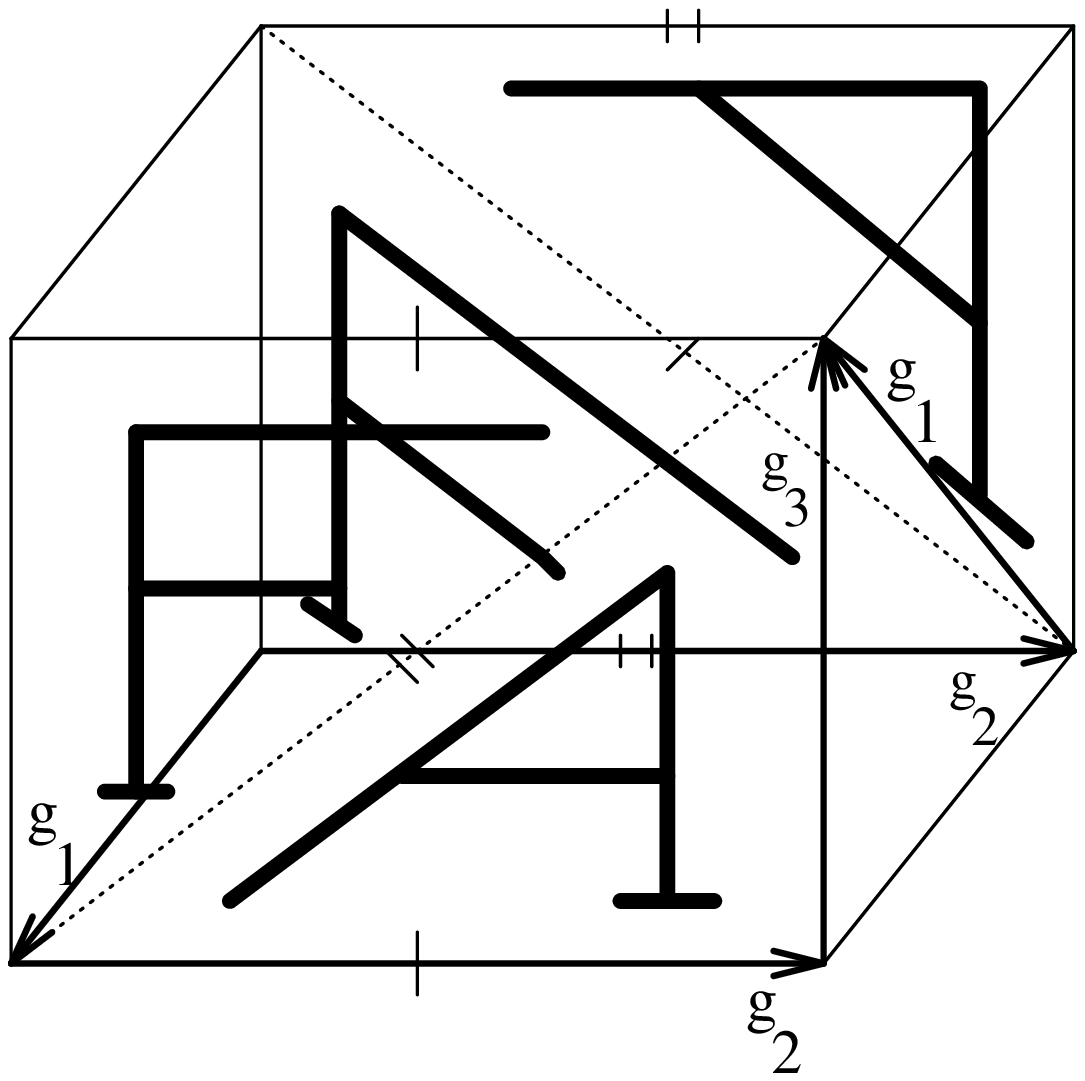}
  \end{center}
\vspace{.3cm}

\begin{flushright}
  \parbox{12cm}{\small Fig.1. The topology of b/1. Letters A and F show how
    the front and back sides are identified. The right and left sides,
    and the top and bottom sides are identified normally.
    The five arrows shows actions of $g_i$'s,
    illustrating the first relation in Eq.(12), $g_3g_2g_1=g_1g_2$.}
\end{flushright}
\end{figure}

We then consider the possible conjugations by $\Isomf={\rm Nil}$.  For
the conjugation of $g_i$'s by $h=({h^1},{h^2},{h^3})\in{\rm Nil}$ (For
typographical convenience, we sometimes write components of group
horizontally), we have
\begin{equation}
        hg_1h\inv=\vector{\dug 11}{\dug 12}{\dug 13+h^1\dug 12-h^2\dug 11},\,
        hg_2h\inv=\vector{\dug 21}{\dug 22}{\dug 23+h^1\dug 22-h^2\dug 21},\,
        hg_3h\inv=g_3.
        \label{b2-6}
\end{equation}
We can make the third components of $g_1$ and $g_2$ zero if we take $h$
as $h^1=(\dug13\dug21-\dug11\dug23)/\Lapp_3{}^3$,
$h^2=(\dug13\dug22-\dug12\dug23)/\Lapp_3{}^3$.  After all, our
representation of $\pi_1$ in Nil reduces to
\begin{equation}
        g_1=\vector{\dug11}{\dug12}{0},\,
        g_2=\vector{\dug21}{\dug22}{0},\,
        g_3=\vector{0}{0}{\Lapp_3{}^3}.
        \label{b2-7}
\end{equation}
The nonvanishing four independent components in these $g_i$'s determine
the initial identifications in the universal cover with metric
\reff{b2-2'}.

To proceed further calculations, we here cite the definition given in \I\
of the \Teich parameters for b/1 and some related properties.  We denote
the standard universal cover as $(\R^3,\ustd ab)$, where the standard
metric $\ustd ab$ is given by (Eq.(75) in \I)
\begin{equation}
        \d l^2=\d x{}^2+\d y{}^2+(\d z-x\d y)^2.
        \label{b2-8}
\end{equation}
Any compact homogeneous 3-manifold classified in b/1 is globally
conformally isometric to manifold $(\R^3,\ustd ab)/A$, where $A$ is a
covering group whose generators are given by
\begin{equation}
        a_1=\vector{\a11}00,\,
        a_2=\vector{\a21}{\a22}0,\,
        a_3=\vector 00{\a11\a22}.
        \label{b2-13}
\end{equation}
Then, the \Teich parameters are $\tau=(\a11,\a21,\a22)$ (Eq.(129) in
\I).  We can see that the map
\begin{equation}
        s_\theta:\,
        \vector{x}{y}{z} \goes
        \svector{R_\theta\svector{x}{y}}{z+\zeta_\theta(x,y)}
        \label{b2-10}
\end{equation}
is a 1-parameter isometry for $(\R^3,\ustd ab)$, where $R_\theta$ is the
rotation matrix by angle $\theta$, and $\zeta_\theta$ is defined by
\begin{equation}
        \zeta_\theta(x,y)\equiv
        \rcp2((x^2-y^2)\cos\theta-2xy\sin\theta)\sin\theta.
        \label{b2-11}
\end{equation}
We here remark that $s_\theta$ is not an element of \Isomf\ but of
$\Isom\uMt$, and therefore $\Isomf\neq\Isom\uMt$ in the b/1 model.  For
an element $h\in$Nil, conjugation by $s_\theta$ is given by
\begin{equation}
        s_\theta\vector{h^1}{h^2}{h^3}s_\theta\inv=
        \svector{R_\theta\svector{h^1}{h^2}}{h^3+\zeta_\theta(h^1,h^2)}.
        \label{b2-12}
\end{equation}

Note that metric \reff{b2-2} is rewritten as
\begin{equation}
        \d l^2={\h1\h2\over\h3}(\d x'{}^2+\d y'{}^2+(\d z'-x'\d y')^2)
        \label{b2-8'}
\end{equation}
with
\begin{equation}
  \vector{x'}{y'}{z'}=
  \vector{\sqrt{\frac{\h3}{\h2}}x}
  {\sqrt{\frac{\h3}{\h1}}y}
  {\frac{\h3}{\sqrt{\h1\h2}}z},
        \label{b2-9}
\end{equation}
where $\h\alpha$ are regarded as constants.  If we view this coordinate
transformation as a diffeomorphism and drop the constant conformal
factor of metric \reff{b2-8'}, the resulting metric coincides with the
standard metric \reff{b2-8}.  This diffeomorphism is obviously an
element of the HPDs \cite{AS,KTH}, from the form of metric \reff{b2-8'},
so that the transformation $(x,y,z)\goes(x',y',z')$ is an (outer-)
automorphism of Nil.  The image of $g_i$'s, which acts on metric
\reff{b2-8'} (or metric \reff{b2-8}), is
\begin{equation}
        g_1'=\vector{\dug11{}'}{\dug12{}'}{0},\,
        g_2'=\vector{\dug21{}'}{\dug22{}'}{0}.
        \label{b2-14}
\end{equation}
Here,
\begin{equation}
        \dug11{}'=\pfaci\dug11,\,\dug12{}'=\pfacii\dug12,
        \dug21{}'=\pfaci\dug21,\,\dug22{}'=\pfacii\dug22.
        \label{b2-15}
\end{equation}
Generator $g_3$ is automatically determined by $g_1$ and $g_2$ (see
Eq.\reff{b2-7}), so we will concentrate on $g_1$, $g_2$ and the images
of them by automorphisms.  Since Eq.\reff{b2-14} is not of the form of
Eq.\reff{b2-13}, it does not yet give the \Teich parameters.  To get
them, we take a conjugation of Eq.\reff{b2-14} by the (full) isometry of
Nil, which is given by Nil itself with $s_\theta$.  We can ``rotate''
the two-dimensional vectors $(\dug i1{}',\dug i2{}')$ $(i=1,2)$ by
conjugations by $s_\theta$ (Eq.\reff{b2-12}), leaving the third
components zero by a conjugation by Nil like the way we obtained
Eq.\reff{b2-7}.  So, we arrive at
\begin{equation}
        a_1=hs_{\theta_1}g_1's_{\theta_1}\inv h\inv=
        \vector{\sqrt{(\dug11{}')^2+(\dug12{}')^2}}{0}{0}
        \label{b2-16}
\end{equation}
and
\begin{equation}
        a_2
        =\vector{ \dug21{}'\cos\theta_1-\dug22{}'\sin\theta_1 }
                { \dug21{}'\sin\theta_1+\dug22{}'\cos\theta_1 }
                { 0 }
        =\rcp{\sqrt{(\dug11{}')^2+(\dug12{}')^2}}
                \vector{\dug11{}'\dug21{}'+\dug12{}'\dug22{}'}
                {\dug11{}'\dug22{}'-\dug12{}'\dug21{}'}{0},
        \label{b2-17}
\end{equation}
where
\begin{equation}
        \cos\theta_1=\frac{\dug11{}'}
                {\sqrt{(\dug11{}')^2+(\dug12{}')^2}},\,
        \sin\theta_1=\frac{-\dug12{}'}
                {\sqrt{(\dug11{}')^2+(\dug12{}')^2}},
        \label{b2-18}
\end{equation}
and $h$ is an element of Nil.  Using Eq.\reff{b2-15}, we obtain the
final form of the \Teich parameters;
\begin{eqnarray}
        \a11\wa\sqrt{\frac{\h3}{\h2}(\dug11)^2+\frac{\h3}{\h1}(\dug12)^2},
                \nonumber \\
        \a21\wa\rcp{\a11}\paren{ {\h3\over\h2}\dug11\dug21
                +{\h3\over\h1}\dug12\dug22 }, \nonumber \\
        \a22\wa{\Lapp_3{}^3\over\a11}\pfaciii.
 \label{b2-19}
\end{eqnarray}

In Eq.\reff{b2-19}, parameters $\dug11,\dug12,\dug21,\dug22$ and hence
$\Lapp_3{}^3$ are constants, and $\h\alpha$'s are functions of $t$.  The
metric components $\dh11,\dh22$, and $\dh33$ are determined by
substituting Eqs.\reff{b2-2'} and \reff{b2-2} into Einstein's equation,
and moreover we must exclude the degrees of freedom of HPDs from them
(See Sec. \ref{degree}). Hence, the number of free parameters that the
metric components can have coincides with the known number of degrees of
freedom of the conventional (open) Bianchi models \cite{KSMH}.  For the
vacuum Bianchi II, the number of free parameters in the metric functions
is two, i.e., $\dim\bar U=2$.  With the four parameters specifying the
initial identifications, $\dim\bar C_u=4$, the total number of
dynamical degrees of freedom of the present vacuum b/1 model is six.
(cf. proposition \ref{prop1-1})
The dynamical variables are the \Teich parameters $\a11,\a21$, and
$\a22$, {\it and} the total volume
\begin{equation}
        v=(\Lapp_3{}^3)^2 \sqrt{\dh11\dh22\dh33}.
        \label{b2-21}
\end{equation}
Remember that the \Teich parameters are defined with respect to the
standard universal cover which is isometric to the universal cover \UCh\ 
up to a global conformal factor.  In fact, it is clear that, if we know
the values of them, we can completely construct the original compact
3-manifold.

\noindent
{\it Additional remarks} :

We end this subsection with some common remarks to the subsequent
subsections, which will be helpful for understanding the rather unique
concept employed throughout this section.  Since as in Sec.\ref{prelim}
any compact locally homogeneous 3-manifold can always be smoothly
specified by some \Teich parameters, curvature parameters, and the
volume, we are regarding them, rather than $\dh\alpha\alpha$'s, as the
dynamical variables of the compact homogeneous universe.  In the b/1
case, they are the four parameters $(\a11,\a21,\a22,v)$, as pointed out.
(There are no curvature parameters in this case).  One remark we want to
emphasize here is that we are on the standpoint that we do {\it not} ask
whether or not such the dynamical variables {\it directly} fulfill some
dynamical differential equations, since we do not need them to obtain
the time-development of the ``dynamical variables''.  (We will however
discuss this problem in a separate work. See also Sec.\ref{conc}.)  Of
course, this is not to say we do not need Einstein's equation.  The role
of Einstein's equation in our calculation is to fix the universal cover,
i.e. to fix $\dh\alpha\alpha$'s.

One of the main conclusions in this subsection was that once the
universal cover is set fixed by Einstein's equation, we automatically
get the time-development of the dynamical variables through
Eqs.\reff{b2-19} and \reff{b2-21}.  In this sense, Eqs.\reff{b2-19} and
\reff{b2-21} are the {\it kinematical} relation between the universal
cover and the dynamical variables. Another remark is therefore the fact
that even if some matter fields are included and the form of
$\dh\alpha\alpha$'s accordingly vary, Eqs.\reff{b2-19} and \reff{b2-21}
are invariant.  Correspondingly, while the degrees of freedom, $\dim\bar
U$, of the universal cover may vary, those, $\dim\bar C_u$, of the
initial identifications are invariant.

\subsection{The f1/1($n$) model: the compact Bianchi VI${}_0$ model \label{f}}

The multiplication rule of Sol (=Bianchi \sz\ group) is given by;
\begin{equation}
        \vector{g^1}{g^2}{g^3}\vector{h^1}{h^2}{h^3}=
        \vector{g^1+e^{-g^3}h^1}{g^2+e^{g^3}h^2}{g^3+h^3},
        \label{b6-1}
\end{equation}
where $g,h\in {\rm Sol}$.  We can easily check that the 1-forms
\begin{equation}
        \s1=\rcp{\sqrt2}(e^z\d x+e^{-z}\d y),\,
        \s2=\rcp{\sqrt2}(-e^z\d x+e^{-z}\d y),\,
        \s3=\d z
        \label{b6-1.2}
\end{equation}
are invariant under the left action of Sol if $(x,y,z)$ is identified
with an element of Sol.  We therefore have the following invariant
metric
\begin{equation}
  \d l^2=\h1(\s1)^2+\h2(\s2)^2+\h3(\s3)^2.
        \label{b6-2}
\end{equation}
As in the case of Bianchi II, the four dimensional universal cover
metric is Eq.\reff{b2-2'} with the above $\d l^2$.  In contrast to the
Bianchi II case, the isometries and the extendible isometries coincide;
$\Isom\uMt=\Isomf=$ (Sol plus three discrete elements).  For future use,
we present one of the three discrete elements here.  It is
\begin{equation}
        h: (x,y,z)\goes(-x,-y,z).
        \label{b6-7}
\end{equation}

The fundamental group $\pi_1$ of a compact manifold modeled on Bianchi
\sz\ geometry is given by (See Eq.(145) in \I)
\begin{equation}
        \pi_1=\angl{g_1,g_2,g_3;[g_1,g_2],g_3g_1g_3\inv g_2\inv,
        g_3g_2g_3\inv g_1g_2^{-n}},
        \label{b6-3}
\end{equation}
where $\abs n>2$.  Because of the coincidence of the two isometry
groups, we need not do new calculations to find embeddings of $\pi_1$ in
\Isomf\ other than those shown in \I.  We simply show the results.

Let
\begin{equation}
        c_3\equiv\ln \frac{\abs{n+\sqrt{n^2-4}}}{2}.
        \label{b6-4}
\end{equation}
If $n>2$, then $e^{-c_3}$ and $e^{c_3}$ are the eigenvalues of matrix
$\smatrix 01{-1}n$, and so are $-e^{-c_3}$ and $-e^{c_3}$, if $n<-2$.
Let $(u_1,v_1)$ and $(u_2,v_2)$ be the normalized eigenvectors
corresponding to the two eigenvalues, i.e.,
\def\sptmp{\sqrt{\abs{n+\sqrt{n^2-4}}\over2}}
\def\smtmp{\sqrt{\abs{n-\sqrt{n^2-4}}\over2}}
\begin{equation}
        \svector{u_1}{v_1}=\rcp{\sqrt{\abs n}}\svector{\sptmp}{\smtmp},\;
        \svector{u_2}{v_2}=\rcp{\sqrt{\abs n}}\svector{\smtmp}{\sptmp}.
        \label{b6-4'}
\end{equation}
Then we can embed the generators of $\pi_1$ in \Isomf\ as (See Eqs.(156)
and (157) in \I)
\begin{equation}
        g_1=\vector{\alpha_0u_1}{\alpha_0u_2}{0},\;
        g_2=\vector{\alpha_0v_1}{\alpha_0v_2}{0},\;
        g_3=\vector{0}{0}{c_3}
        \label{b6-5}
\end{equation}
for $n>2$, or
\begin{equation}
        g_1=\vector{\alpha_0u_1}{\alpha_0u_2}{0},\;
        g_2=\vector{\alpha_0v_1}{\alpha_0v_2}{0},\;
        g_3=h\circ\vector{0}{0}{c_3}
        \label{b6-6}
\end{equation}
for $n<-2$, where $h$ is defined in Eq.\reff{b6-7}.  We thus find that
the parameter for the initial identifications is only $\alpha_0$ in
Eq.\reff{b6-5} or Eq.\reff{b6-6}.

Before giving the time-development of the \Teich parameters, we take
this opportunity to present a pictorial account to the topology of a
compact Sol: f/1(\n).  Manifold f/1(\n) is a torus-bundle over $S^1$.
The relation \reff{b6-3} implies that $g_1$ and $g_2$ generate the fiber
torus.  We can understand the topology of f/1(\n) by observing the
gluing map generated by $g_3$ which maps generators of a torus to
another generators of a torus.  From the relation \reff{b6-3}, we
observe that \def\tg{{\tilde g}}
\begin{equation}
        \tg_1\equiv g_3g_1g_3\inv=g_2,\;
        \tg_2\equiv g_3g_2g_3\inv=g_2^ng_1\inv.
        \label{b6-7'}
\end{equation}
This means that $g_3$ maps a parallelogram spanned by $g_1$ and $g_2$ on
an $x$-$y$ plane to another parallelogram spanned by $\tg_1$ and $\tg_2$
on another $x$-$y$ plane (Fig.2 (A) and (B)).  If we ``cut'' and
translate the second parallelogram by the actions of $g_1$ and $g_2$,
then we can take a fundamental region as a parallelopiped of which the
bottom and top surfaces are spanned by $g_1$ and $g_2$ (Fig.2 (B) and
(C)).  When identifying the bottom surface to the top, the surface is
stretched by $n$ times, and then folded.  (A geodesic congruence along
$z$-axis in f/1($n$) will therefore behave extremely chaotically after a
journey over some periods.)

\begin{figure}[btp]
\begin{center}
    \leavevmode
\epsfysize=10cm
\epsfbox{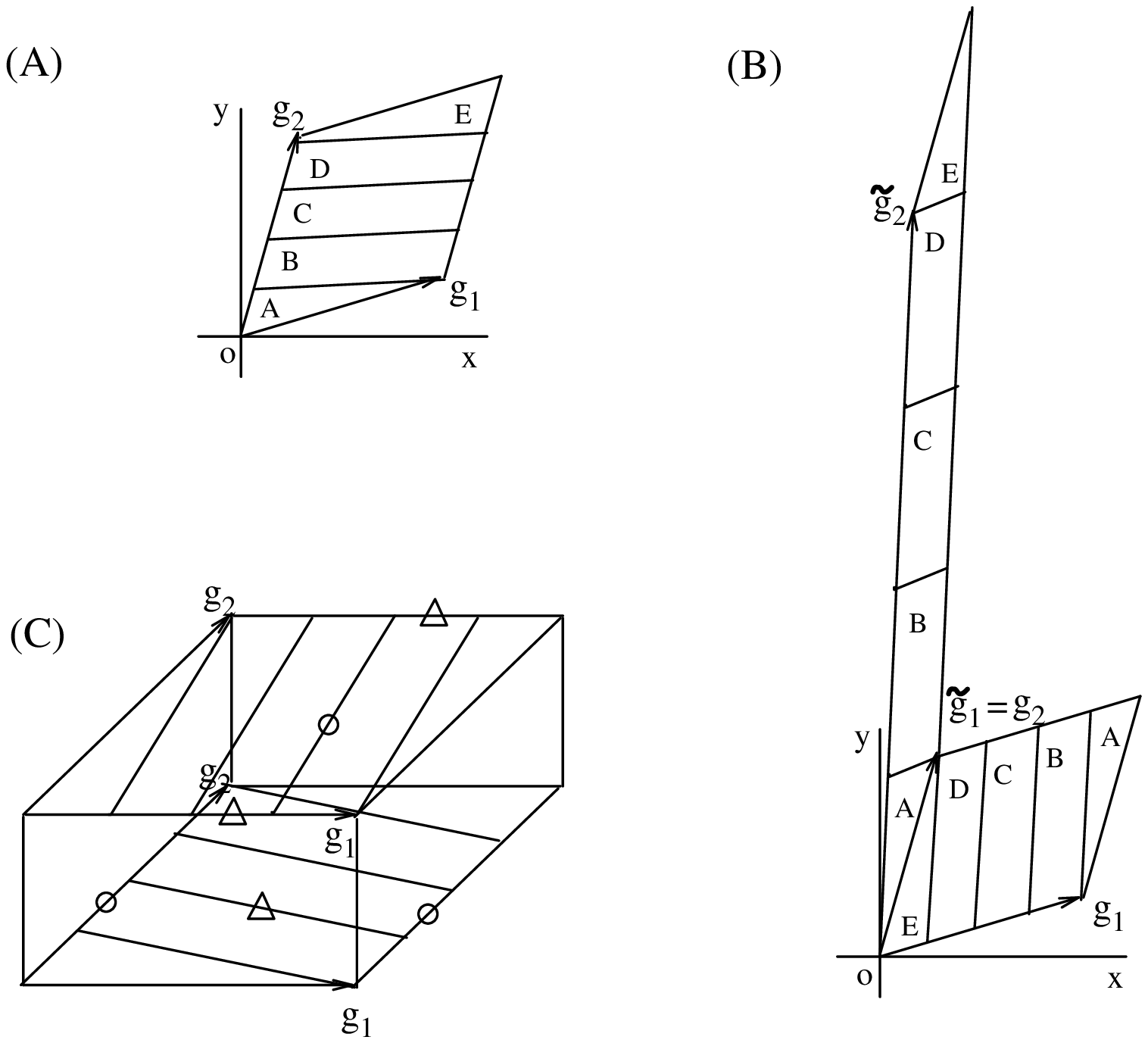}
\end{center}
\vspace{.8cm}

\begin{flushright}
  \parbox{12cm}{\small Fig.2: The topology of f/1($\n=4$).
  $g_1$ and $g_2$ span a parallelogram in a $x$-$y$ ,say $z=z_0$, plane
  (Fig.(A)).
  The stretched parallelogram, shown in Fig.(B), spanned by
  $\tg_1=g_2$ and $\tg_2=g_2^4g_1\inv$ in $z=z_0+c_3$ plane
  can be identified by the actions of $g_1$, $g_1g_2\inv$, ..., $g_1g_2^{-4}$
  with the parallelogram spanned by $g_1$ and $g_2$ in a way that
  letters A$\sim$E in Fig.(B) label the corresponding regions.
  The corresponding regions in Fig.(A) are labeled by the same letters.
  So, the up and down sides of the parallelopiped in Fig.(C), which shows
  a fundamental region of f1/1(4), are identified as indicated
  by circles and triangles.
  The front and back sides, and the right and left sides are
  identified in the trivial way.
  }
\end{flushright}
\end{figure}

Let us return to the operation to identify the \Teich parameter.  Note
that we can transform the spatial metric \reff{b6-2} into
\begin{equation}
  \label{b6-8}
  \d l^2= \h3\bra{\rcp2\sqrt{\frac{\h1}{\h2}}
    \paren{e^{z'}\d x'+e^{-z'}\d y'}^2+
  \rcp2\sqrt{\frac{\h2}{\h1}}\paren{-e^{z'}\d x'+e^{-z'}\d y'}^2+\d z'{}^2 },
\end{equation}
where
\begin{equation}
  \label{b6-9}
  \vector{x'}{y'}{z'}=\vector{\frac{(\h1\h2)^{1/4}}{\sqrt{\h3}}x}
  {\frac{(\h1\h2)^{1/4}}{\sqrt{\h3}}y}{z}.
\end{equation}
The transformation $(x,y,z)\goes(x',y',z')$ defined by Eq.\reff{b6-9} is
an automorphism of Sol.  So, the appropriate action of identifications
on $(x',y',z')$ is given by
\begin{equation}
  \label{b6-10}
  g_1'=\frac{(\h1\h2)^{1/4}}{\sqrt{\h3}}
  \vector{\alpha_0u_1}{\alpha_0u_2}{0},\,
  g_2'=\frac{(\h1\h2)^{1/4}}{\sqrt{\h3}}
  \vector{\alpha_0v_1}{\alpha_0v_2}{0},\,
  g_3'= \vector{0}{0}{c_3}
\end{equation}
for $n>2$, or $g_3'= h\circ({0},{0},{c_3})$ for $n<-2$.  This already
coincides with the parametrization of the \Teich space, i.e.  the only
\Teich parameter is
\begin{equation}
  \label{b6-11}
  \alpha(t)=\frac{(\h1\h2)^{1/4}}{\sqrt{\h3}}\alpha_0.
\end{equation}
It is worth noting that we can observe from Eq.\reff{b6-10} that the
\Teich deformations of f1/1(n) are the variations of the ratio of the
area of the fiber torus to the length of the base $S^1$.

The dynamical variables (in configuration space) are the \Teich
parameter $\alpha$, the 3-volume $v=\h1\h2\h3(\alpha_0)^4$ up to a
function depending on $n$, and the curvature control parameter
$\lambda=\ln(\h1/\h2)$.  The number of dynamical degrees of freedom is
four; one is for $\alpha_0$, and three is for the parameters contained
in the four-dimensional universal cover.

\subsection{The a1/1 model: a compact model on Bianchi VII${}_0$ \label{a1}}

The multiplication rule of Bianchi \svz\ group is given by;
\begin{equation}
        \vector{g^1}{g^2}{g^3}\vector{h^1}{h^2}{h^3}=
        \svector{\svector{g^1}{g^2}+R_{g^3}\svector{h^1}{h^2}}{g^3+h^3},
        \label{b7-1}
\end{equation}
where $g,h\in$ Bianchi \svz\ group, and $R_{g^3}$ is the rotation matrix
by angle $g^3$.  The 1-forms
\begin{equation}
        \s1=\cos z\d x+\sin z\d y,\,
        \s2=-\sin z\d x+\cos z\d y,\,
        \s3=\d z
        \label{b7-1.2}
\end{equation}
are invariant under the left action of Bianchi \svz.  Hence the
invariant metric is
\begin{equation}
  \d l^2=\h1(\s1)^2+\h2(\s2)^2+\h3(\s3)^2.
  \label{b7-2}
\end{equation}
As usual, the four dimensional universal cover metric is Eq.\reff{b2-2'}
with the above $\d l^2$.  Since, as in the Bianchi \sz\ case in the
previous subsection, the isometries and the extendible isometries of
\UCh\ coincide, $\Isom\uMt=\Isomf=$ (Bianchi \svz\ group plus three
discrete elements), our calculations to do will be similar to those in
the Bianchi \sz\ case.

The compact 3-manifold we consider here is a1/1, which is homeomorphic
to the 3-torus $T^3$.  The three generators, $g_1,g_2$ and $g_3$, of the
fundamental group of a1/1 are hence all commutative.

The embedding of the fundamental group in \Isomf\ up to conjugacies by
$\Isomf(=\Isom\uMt)$ is already given in \I, which reads
\begin{equation}
  \label{b7-3}
  g_1=\vector{\dug11}{0}{2l\pi},\,
  g_2=\vector{\dug21}{\dug22}{2m\pi},\,
  g_3=\vector{\dug31}{\dug32}{2n\pi},
\end{equation}
where $l,m,n$ are integers.

To obtain the time-development of the \Teich parameters, we, as usual,
first note that we can transform the spatial metric \reff{b7-2} into
\begin{equation}
  \d l^2= \h3\paren{ \sqrt{\frac{\h1}{\h2}}(\cos z'\d x'+\sin z'\d y')^2+
    \sqrt{\frac{\h2}{\h1}}(-\sin z'\d x'+\cos z'\d y')^2
  +\d z'{}^2 },
  \label{b7-4}
\end{equation}
where
\begin{equation}
  \label{b7-5}
  \vector{x'}{y'}{z'}=\vector{\frac{(\h1\h2)^{1/4}}{\sqrt{\h3}}x}
  {\frac{(\h1\h2)^{1/4}}{\sqrt{\h3}}y}{z}.
\end{equation}
Here, the metric \reff{b7-4} coincides with the standard metric, given
in \I, of Bianchi \svz\ up to global conformal factor.

Since the transformation $(x,y,z)\goes(x',y',z')$ is an automorphism of
the Bianchi \svz\ group, we can easily obtain the actions on the
standard metric \reff{b7-4} of the generators \reff{b7-3}.  We
immediately get
\begin{equation}
  \label{b7-6}
  a_1=\vector{\a11}{0}{2l\pi},\,
  a_2=\vector{\a21}{\a22}{2m\pi},\,
  a_3=\vector{\a31}{\a32}{2n\pi},
\end{equation}
where
\begin{equation}
  \label{b7-7}
  \a ij=\frac{(\h1\h2)^{1/4}}{\sqrt{\h3}}\dug ij,\, 
  (i,j)=(1,1),(2,1),(2,2),(3,1),{\rm and}\, (3,2).
\end{equation}
Here, $a_i$'s are the images of $g_i$'s by the automorphism, and
$\a11\sim\a32$ are the \Teich parameters.

To summarize, the five constants $\dug11\sim\dug32$ determine the
initial identifications on an initial surface in \UC.  The universal
cover \UC\ have three arbitrary parameters in vacuum, and hence the
number of the dynamical degrees of freedom is 8 $(=5+3)$.  We have seven
dynamical variables; the five \Teich parameters, the curvature control
parameter $\lambda=\ln(\h1/\h2)$, and the 3-volume
$v=\sqrt{\h1\h2\h3}\det(g_1,g_2,g_3)$.

\subsection{The a2/1 model: a compact model on Bianchi I \label{b2}}

In this subsection, we give the time-development of the \Teich
parameters of the b2/1 model ($\simeq T^3$) on vacuum Bianchi I, by a
method other than the one finding an automorphism explicitly.  This is
done by calculating invariants under the automorphisms, like lengths of
minimal loops and angles between the loops of the compact homogeneous
manifold.  To this end, we introduce a matrix representing such
invariants as follows.

Let $\Gam$ be a covering group acting on $(\Mtil,\tilde h_{ab})$.  For
$a\in\Gam,\, p\in\Mtil$, let the map $\gamma_{a,p}:\, \R\goes\Mtil$ be
the {\it geodesic} satisfying
\begin{eqnarray}
        \gamma_{a,p}(0)\wa p, \nonumber \\
        \gamma_{a,p}(1)\wa a(p).
        \label{b1-2}
\end{eqnarray}
Then, we define the map $\upsilon_p:\, \Gam\goes V_p$ by relating \Gam\ 
to the geodesic generator at $p$;
\begin{equation}
        \upsilon_p(a)=
        \left.{\d\gamma_{a,p}(\lambda)\over\d\lambda}\right|_{\lambda=0}.
        \label{b1-4}
\end{equation}
Finally, let all the independent generators of \Gam\ be $a_i$
$(i=1,\cdots,n)$.  Then the {\it loop matrix} defined by
\begin{equation}
        H_{ij}(p)=\tilde h(\upsilon_p(a_i),\upsilon_p(a_j))
        \label{b1-5}
\end{equation}
will contain all the information concerning the global geometry of
$M=\Mtil/\Gam$.  Here, we have dropped the abstract indices of the
metric and the vectors in the r.h.s..  It is worth noting that the
$p$-dependence of $H_{ij}(p)$ decides whether $M$ is locally homogeneous
or globally homogeneous.  That is, if $H_{ij}(p)$ is independent of $p$,
then $M$ is globally homogeneous, and if not so, then homogeneity of $M$
is local.

We are in a position to calculate the time-development of the a2/1 model
in vacuum.  Our four dimensional universal cover is the Kasner solution
$=(\R^4,\udg ab)$, where with the usual coordinates $(t,x,y,z)$, $\udg
ab$ is given by Eq.\reff{mc5} with $\sigma\equiv
p_1+p_2+p_3=1=(p_1)^2+(p_2)^2+(p_3)^2$.  Each homogeneous spatial
section is given by $(\R^3,\udh ab)$ with $\udh ab$ being Eq.\reff{mc7}.
The covering group may be generated by three commuting generators, for
which we write as
\begin{equation}
        g_i=\vector{\dug i1}{\dug i2}{\dug i3},\, (i=1\sim 3).
        \label{b1-6}
\end{equation}
Here, the action of $g_i$ on $p=({x_0},{y_0},{z_0})$ on $(\R^3,\udh ab)$
is given by
\begin{equation}
        g_i\vector{x_0}{y_0}{z_0}=
        \vector{\dug i1+x_0}{\dug i2+y_0}{\dug i3+z_0}.
        \label{b1-7}
\end{equation}
Since the extendible isometry group of each slice is isomorphic to the
commutative group $\R^3$, we see that no nontrivial conjugation occurs.
This implies that we cannot simplify the components of $g_i$'s more than
the original form of Eq.\reff{b1-6}. 

We can at this point count the dynamical degrees of freedom of the
present model. Immediately can we see that the Kasner parameter carries
the part of dynamical degrees of freedom in the universal cover,
$\dim\bar U=1$, while $\dug i\alpha$'s in Eq.\reff{b1-6} carry the
part in the initial identifications, $\dim\bar C_u=9$. The total
dynamical degrees of freedom is therefore 10($=\dim\bar U+\dim\bar
C_u$). (cf. proposition \ref{prop1-1})

Now, return back to the procedure to get the \Teich parameters.
We can easily find the geodesics
satisfying Eq.\reff{b1-2}, and then get the generator at $p$ as follows.
\begin{equation}
        \gamma_{g_i,p}(\lambda)=
        \vector{\dug i1\lambda+x_0}{\dug i2\lambda+y_0}{\dug i3\lambda+z_0},\,
        \upsilon_p(g_i)=\vector{\dug i1}{\dug i2}{\dug i3}.
        \label{b1-8}
\end{equation}
From this, we have
\begin{eqnarray}
        H_{ij}(p)\wa\tilde h(\upsilon_p(g_i),\upsilon_p(g_j)) \nonumber \\
        \wa \sum_{\alpha=1}^3t^{2p_\alpha}\dug i\alpha\dug j\alpha.
        \label{b1-10}
\end{eqnarray}

On the other hand, any flat $T^3$ (a2/1) can be implemented in the
standard Euclid metric
\begin{equation}
        \d l^2=\d x^2+\d y^2+\d z^2
        \label{b1-11}
\end{equation}
with six \Teich parameters \cite{tei} in three generators
\begin{equation}
        a_1=\vector{\a11}{0}{0},\,
        a_2=\vector{\a21}{\a22}{0},\,
        a_3=\vector{\a31}{\a32}{\a33}.
        \label{b1-12}
\end{equation}
Components of $\upsilon_p(a_i)$ is the same as Eq.\reff{b1-8} with $\dug
ij$ replaced by $\a ij$ with $\a12=\a13=\a23=0$.  Using Eq.\reff{b1-11}
as $\tilde h$, we have
\begin{eqnarray}
        H_{ij}\wa\a i1\a j1+\a i2\a j2+\a i3\a j3 \nonumber \\
        \wa\matrix{(\a11)^2}{\a11\a21}{\a11\a31}
                {}{(\a21)^2+(\a22)^2}{\a21\a31+\a22\a32}
                {(sym.)}{}{(\a31)^2+(\a32)^2+(\a33)^2}
        \label{b1-13}
\end{eqnarray}

We set equal the two expressions Eqs.\reff{b1-13} and \reff{b1-10} to
get $\a ij$ as time functions with initial parameters $\dug i\alpha$.
Elementary calculations lead to the following results;
\begin{eqnarray}
        (\a 11)^2\wa\sum_\alpha t^{2p_\alpha}(\dug 1\alpha)^2,\,
        \a21=(\sum_\alpha t^{2p_\alpha}\dug 1\alpha\dug 2\alpha)/\a11,\,
        \a31=(\sum_\alpha t^{2p_\alpha}\dug 1\alpha\dug 3\alpha)/\a11,
         \nonumber \\
        (\a22)^2\wa\Lap^2/(\a11)^2,\,
        \a32=-\a22(\sum_\alpha t^{2(\sigma-p_\alpha)}
                \Lapp_3{}^\alpha\Lapp_2{}^\alpha)/\Lap^2,\,
        \a33=(\det g)^2t^{2\sigma}/\Lap^2,
        \label{b1-14}
\end{eqnarray}
where $\Lap^2\equiv\sum_\alpha
t^{2(\sigma-p_\alpha)}(\Lap_3{}^\alpha)^2$, $\sigma\equiv\sum_\alpha
p_\alpha=1$, and $\Lapp_i{}^\alpha{}$ is the $(i,\alpha)$th cofactor of
the matrix $(\dug i\alpha)$.  It would be useful, especially in getting
$\a33$, to note that the determinant of $H_{ij}$ is given by
\begin{equation}
        \det H=(\det g)^2t^{2\sigma}=(\a11)^2(\a22)^2(\a33)^2.
        \label{b1-15}
\end{equation}

\section{Conclusions}
\label{conc}

We have given a general method of construction of compact homogeneous
universes.  This is accomplished by taking identifications in a
universal cover \UC.  The universal cover must satisfy Einstein's
equation, and the degrees of freedom of all the possible \diffeos\ must
be subtracted.  The identifications in the universal cover are
implemented by a discrete subgroup of the extendible isometries, \Isomf.
At this stage, one takes the conjugations by \Isomf, and finally we can
obtain a system of compact homogeneous universes which is free from any
\diffeos, i.e., the free parameters in the metric and in the
identifications are the dynamical degrees of freedom of the system.
This method of construction is evidently applicable to any system of
compact homogeneous universes (i.e., compact models on the Bianchi class
A, class B, and the Kantowski-Sachs-Nariai models).

We have considered the dynamical variables of the system to be
parameters specifying spatial sections completely.  In this sense, the
\Teich parameters are dynamical variables, as well as the 3-volume and
the possible curvature parameters.  It is important to note that the
number, $f$, of dynamical degrees of freedom is less than double the
number, $d$, of dynamical variables.  As we have seen in the explicit
examples, not all of the initial velocities can be arbitrarily chosen.
In some cases, $f$ is less than $2d$ by 2.  This could be explained by
the Hamiltonian constraint.  In the other cases, however, $f$ is less
than $2d-2$.  These could be well understood if we study whether the
dynamical system admits a canonical structure.  This is also needed in
canonical quantization of compact homogeneous universes.  This problem
will be discussed in a separate work.

Although we focused on the time-developments of the \Teich parameters
and the dynamical degrees of freedom, our framework of compact
homogeneous universes should be useful in wide variety of problems in
astrophysics, observational cosmology, fundamental problems of
relativity, quantum cosmology, and quantum gravity.  For example, the
behavior of geodesics in a compact universe can become chaotic (cf.
Sec.\ref{f}), which fact may provide an interesting cosmological model.
The problem of strong cosmic censorship for compact homogeneous
universes is also of great interest, which is being investigated by some
groups (See, e.g., \cite{Cru}).

\section*{Acknowledgments}
M. T. thanks Soryushi Shogakukai for financial support.  T. K.
acknowledges financial support from the Japan Society for the Promotion
of Science and the Ministry of Education, Science and Culture.  This
work is patially supported by the Grant-in-Aid for Scientific Research
of the Ministry of Education, Science, and Culture of Japan
(No.02640232)(A.H.).


\begin{thebibliography}{99}
\bibitem{er} For type d (, which correspond to Bianchi VIII),
  the representative metric (or standard metric, see Sec.\ref{cl})
  is incorrect, due to incorrectness of the appendix.
  For detail, see the erratum, which will appear in the journal.
\bibitem{ex} {Here, Eq.\reff{mc3} is understood. The extrinsic curvature
    therefore has the same homogeneity as the spatial metric.}
\bibitem{id} {For simplicity, we identify the set $\bar U$ with a set of
        representatives. Similar identifications are understood also
        for the sets of equivalence classes defined subsequently.}
\bibitem{tei} For convenience, we here do not distinguish the 3-volume
from the \Teich parameters. If do so, we need to regard, e.g., $\a33$
as a function of other \Teich parameters and the 3-volume.
\end{thebibliography}
\end{document}